\documentclass[%
 aip,
cp,  
 amsmath,amssymb,
 reprint,%
]{revtex4-2}

\usepackage{booktabs}
\usepackage{subcaption}
\usepackage{graphicx}
\usepackage{dcolumn}
\usepackage{bm}
\usepackage{tikz}
\usepackage{caption}
\usepackage{subcaption}

\DeclareCaptionFont{ninept}{\fontsize{9}{11}\selectfont}
\DeclareCaptionLabelFormat{figlabel}{\textbf{FIGURE #2.}}
\DeclareCaptionLabelFormat{tablabel}{\textbf{TABLE #2.}}

\usepackage[utf8]{inputenc}
\usepackage[T1]{fontenc}
\usepackage{mathptmx} 

\begin{document}

\title{An Optical Emulation of the BB84 Quantum Key Distribution Protocol}

\author{Alec L. Riso} 
 \email[Corresponding author: ]{alr328@cornell.edu}
 \affiliation{\mbox{
  Department of Physics, Cornell University, Ithaca, New York 14853, USA
}}
\author{Karthik Thyagarajan}%
 \email{kthyagar@purdue.edu}
\affiliation{\mbox{Department of Computer Science, Purdue University, West Lafayette, Indiana 47907, USA
}}

\author{Connor Whiting}
 \email{gcw3@illinois.edu}
\affiliation{\mbox{Department of Physics, University of Illinois Urbana-Champaign, Urbana, Illinois 61801, USA
}}

\author{Katherine Jimenez}
 \email{katiejim@umich.edu}
\affiliation{\mbox{Department of Physics, University of Michigan, Ann Arbor, Michigan 48109, USA}
}

\date{\today} 
\begin{abstract}
Quantum key distribution (QKD) is an approach to secure communication that uses the principles of quantum mechanics to generate a shared secret key which can be used in encrypting messages. Unlike traditional cryptography, which relies on mathematical problems too computationally difficult for a classical computer to solve, an ideal QKD protocol can be secure even against the capabilities of quantum algorithms. The BB84 protocol, a famous implementation of QKD, uses the polarization states of photons to carry information. We constructed an accessible, fiber-optic-based apparatus that emulates the generation of a key through the BB84 protocol. Although a true demonstration of BB84 would require a single-photon regime, our many-photon system shows the information encoding, transmission, and basis-sifting steps of the protocol. Due to its accessibility and visual clarity, our apparatus is intended as an educational tool that makes each logical step of the protocol visible.
\end{abstract}
\vspace{-1em}
\maketitle

\section{\label{sec:level1}Introduction}

Quantum computing has the potential to completely reshape our technological landscape. With applications in areas such as AI, finance, and healthcare, efficient quantum algorithms can give us the ability to solve problems that are currently intractable. However, future advancements in quantum computing can present some alarming consequences. One of the most concerning possibilities is the breaking of asymmetric data encryption methods which keep our money safe and our personal data private. This leaves us vulnerable to attacks on financial data, medical data, and other personally identifying information.

Many of these concerns arise from \textit{Shor’s algorithm} \cite{ShorsPaper}, a quantum algorithm developed by Peter Shor in 1994. It allows for the factorization of numbers to be done in polynomial time with respect to the number of input bits, rather than the subexponential time required by the best-known classical algorithms, such as the general number field sieve\cite{NumberFieldSieve}. As the numbers get larger, the gap between Shor’s algorithm and its classical alternatives widens, potentially shortening solution times from centuries to minutes. Through its application to factorization and discrete logarithm computing problems\cite{DiscreteLogarithms,EllipticCurves}, Shor’s algorithm can effectively break much of our current data encryption schemes, such as RSA, DSA, and ECC. Fortunately, implementing Shor’s algorithm with large numbers would require a quantum computer with significantly more qubits than we currently possess. Although today's largest quantum systems are surpassing 1,000 physical qubits, breaking RSA encryption at current qubit error rates would likely require hundreds of thousands \cite{QubitsToBreakRSA}. However, the pace of development in quantum hardware and quantum error correction suggests this number could keep decreasing.

Even though current quantum systems are too small to accurately run Shor’s algorithm at scale, the possibility still poses a threat. A strategy known as “harvest now, decrypt later” can be employed by bad actors: they collect encrypted data now and store it until a sufficiently powerful quantum computer becomes available. While some recovered information may become obsolete, long-term government operations, financial history, or personal medical records could still be highly damaging if exposed.

One promising direction to mitigate this risk is quantum key distribution (QKD). Like traditional methods, QKD assumes key-based encryption, but it leverages the principles of quantum mechanics to achieve security \cite{QKDOverview}. Two core principles enable QKD's effectiveness. First, quantum states can exist in superposition, a probabilistic mix of possible measurable outcomes. Observation of a quantum state in superposition collapses the state onto one of these possible outcomes, which has the potential to expose the presence of an eavesdropper.  In QKD, an eavesdropper is not certain of the quantum states they intercept, which means they cannot be sure if their measurements are collapsing a superposition and disturbing the state. Second, the no-cloning theorem states that it is impossible to create perfect copies of an arbitrary quantum state \cite{NoCloningTheorem}. This means the eavesdropper cannot copy the quantum states they intercept and simply measure the cloned states. 

These properties make QKD well suited for applications requiring high levels of security, including the secure transmission of classified government and military information, the protection of sensitive financial data and transactions, and the safeguarding of confidential healthcare records. Despite its promise, QKD faces practical challenges. Current QKD systems are expensive and require specialized equipment and expertise, quantum signals degrade over long distances, and integrating QKD with existing infrastructure remains difficult. Nevertheless, research and development in QKD are rapidly advancing. As threats to classical encryption schemes grow, QKD stands out as a promising approach to secure communication.

\subsection{\label{sec:level2}Algorithm Overview}

The fundamental goal of QKD is for two parties to establish a shared secret key, represented by a string of bits. Rather than sending a preexisting key through the channel, QKD distills the key from correlated measurement records that Alice and Bob build up and then compare with each other. The procedure must either keep this key private or, if someone is eavesdropping on the transmission, make that eavesdropping detectable.

This algorithm is done between a transmitter, denoted Alice, and a receiver, denoted Bob. Once the key is securely established and verified, and an encryption scheme is established with this key, Alice can publicly send messages to Bob, which Bob can easily decrypt without worrying that someone else has stolen their data.

Our optical emulation of QKD encodes information in the polarization of light carried by polarization-maintaining fiber-optic cables. Eavesdropping on such a channel is similar to how direct eavesdropping on classical fibers takes place. Much of it occurs at the entry, exit, and transition periods where the encrypted information gets processed or, in the case of the internet, web servers. Directly eavesdropping on the intermediate cable is more difficult and would noticeably disrupt the channel. Other implementations of QKD may be wireless, in which case eavesdropping can come in many other forms. In the case that an eavesdropper is found, another channel may be selected, or the transmission may be attempted at another time.

We focus on the implementation of QKD over fiber optics, modeled after the well-known BB84 protocol \cite{Bennett1984}. We built a communication network that allows real-time emulation of the protocol to viewers. The system showcases how the BB84 protocol works and serves as a valuable educational tool in optics and the potential future of quantum security. It is important to note that this setup uses a large number of photons, and therefore does not exactly replicate the BB84 protocol, which manipulates the quantum states of single photons \cite{Bennett1984}. The setup is modular and combines optics, electronics, and software in a way that makes every step of the algorithm visible and understandable. As the quantum era unfolds, teaching tools like this will be crucial to prepare the next generation of scientists and engineers.

\subsection{\label{sec:level2}Photonic Implementation and Measurement Bases}

In order to emulate quantum key distribution using photonics, we must transport our information using photons. We store this information in the form of the polarization of light emitted from our nanosecond pulse laser. By first polarizing the unpolarized light streaming out from the laser, we can set all photons to a known initial state to modify later.

If two polarizers are oriented exactly perpendicular to each other, a photon that passes through the first polarizer will not pass through the second one. If we think of polarization as a quantum state vector, the vectors are orthogonal in this case. Additionally, two polarizers oriented in the same direction will cause 100\% of the photons that pass through the first polarizer to pass through the second one. These can be thought of as parallel vectors. In our many-photon regime with approximately linear polarization, this interaction can be described classically using Malus’s law: 

\begin{equation}
I = I_0 \cos^2\theta.
\label{eq:malus}
\end{equation}

In this equation, $\theta$ is the angle between the polarization direction of the incident light and the axis of the polarizer, while $I_0$ and $I$ are the incident and transmitted light intensities, respectively. Thus, Malus's law follows the same rules regarding orthogonal and parallel polarization directions. Our system will be in this many-photon regime, but a true BB84 protocol operating with single photons cannot be described classically. Regardless of the regime, we can think of polarization directions as vectors that can form a measurement basis.

Any two orthogonal polarization vectors can form a two-dimensional orthogonal basis. In this context, a basis refers to how we choose to measure a given vector in our two-dimensional polarization space. We can choose any two orthogonal polarization vectors by orienting our polarizer. This forms an orthogonal basis, which can handle any input polarization vector in our two-dimensional space but constrains the measurement output to one of the two basis states, which determines whether or not the light passes through.

Horizontally polarized light is orthogonal to vertically polarized light, and diagonally polarized light is orthogonal to antidiagonally polarized light, thus forming two separate bases as shown in Fig. \ref{phase_diagram}. We can choose to measure light in either basis, but the output must be one of the two specified basis vectors of the basis that we chose. For example, if we measure one photon in the horizontal/vertical (H/V) basis by using a horizontal polarizer, it is forced to choose either the horizontal vector (pass through) or the vertical vector (do not pass through). Thus, if our aim is to get certain results, we would feed in light that is previously polarized to a basis vector (H or V in this example) so that we can fully predict the path the photon will take.

\begin{figure}
\centering
    \begin{tikzpicture}
        \draw[->,ultra thick, black] (0, 0) -- (2, 0) node[anchor=west]{H};
        \draw[->,ultra thick, black] (0, 0) -- (0, 2) node[anchor=south]{V};
        \draw[->,ultra thick, black] (0, 0) -- (-2, 0) node[anchor=east]{H};
        \draw[->,ultra thick, black] (0, 0) -- (0, -2) node[anchor=north]{V};
        \draw[->,ultra thick, red] (0, 0) -- (1.421, 1.421) node[anchor=west]{D};
        \draw[->,ultra thick, red] (0, 0) -- (-1.421, 1.421) node[anchor=east]{AD};
        \draw[->,ultra thick, red] (0, 0) -- (-1.421, -1.421) node[anchor=east]{D};
        \draw[->,ultra thick, red] (0, 0) -- (1.421, -1.421) node[anchor=west]{AD};
        \draw[ultra thick, gray] (0.667, 0) arc (0:45:0.667) node[anchor=west]{45\textdegree};
    \end{tikzpicture}
    \caption{A diagram showing the orientation of polarization states corresponding to two bases.}
    \label{phase_diagram}
\end{figure}

If one measures light polarized as a H/V basis vector in the diagonal/antidiagonal (D/AD) basis (and vice versa), the measurement projects the state onto one of the two D/AD basis vectors rather than reading out a definite preexisting value. Since a given H/V basis vector is 45 degrees from both D/AD basis vectors, a single photon has a 50\% chance of being measured as diagonal and a 50\% chance of being measured as antidiagonal. For the bright, many-photon pulses used in this apparatus, this appears as the optical power dividing approximately equally between the two D/AD outputs, as described by Eq. (\ref{eq:malus}). 

A step-by-step explanation of a BB84-style QKD protocol implemented using the previously described polarization bases is provided in Appendix A.

\section{Methods}

To generate the light, we fed the output of a 980-nm laser directly into a polarization-maintaining (PM) optical fiber. This fiber then passed through an inline polarizer to regularize the polarization of all photons. We use the PM fiber in order to keep the polarization constant after we set the initial state with our inline polarizer. The fiber is then connected to a single-mode optical fiber, which is more vulnerable to polarization change. This fiber first passes through a motorized polarization controller that represents Alice. The Alice controller is manipulated by a servo motor, which changes the polarization state of the photons in one of four ways, depending on the desired bit and the randomly selected basis (H, V, D, or AD). This light travels through another PM fiber, which represents the communication channel. The light then passes into the second polarization controller, which represents Bob. This controller rotates the polarization state of the photons in one of two ways, depending solely on Bob’s selected basis (H/V or D/AD). The servo motor settings are determined by an Arduino UNO, coded with specific degree measures as instructions. 

\begin{figure}
    \centering
    \includegraphics[width=0.7\textwidth]{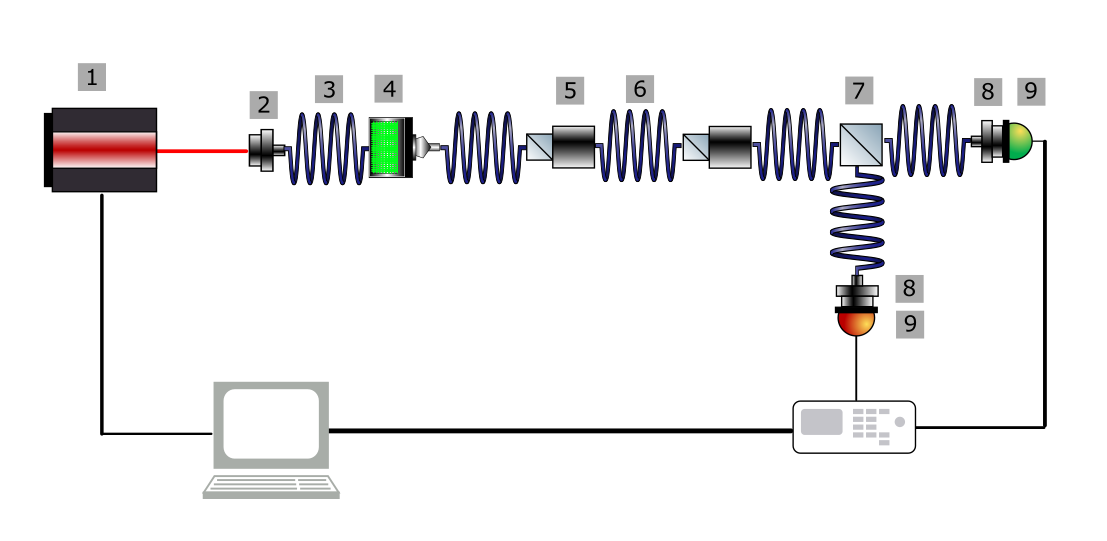}
    \caption{A diagram displaying the optomechanical setup of the physical QKD implementation. Light emerges from a nanosecond pulse laser [1]  and is coupled [2] into a polarization-maintaining fiber-optic cable [3] with an initialized polarization [4]. Alice manipulates the polarization state [5,6] and sends the information to Bob, who manipulates the polarization again. The polarization states are divided by the beamsplitter [7] and collected into one of two photon detectors [8,9]. All part numbers and specifications are detailed in Appendix B.}
    \label{qkd_diagram}
\end{figure}

\begin{figure}
    \centering
    \includegraphics[width=0.5\textwidth]{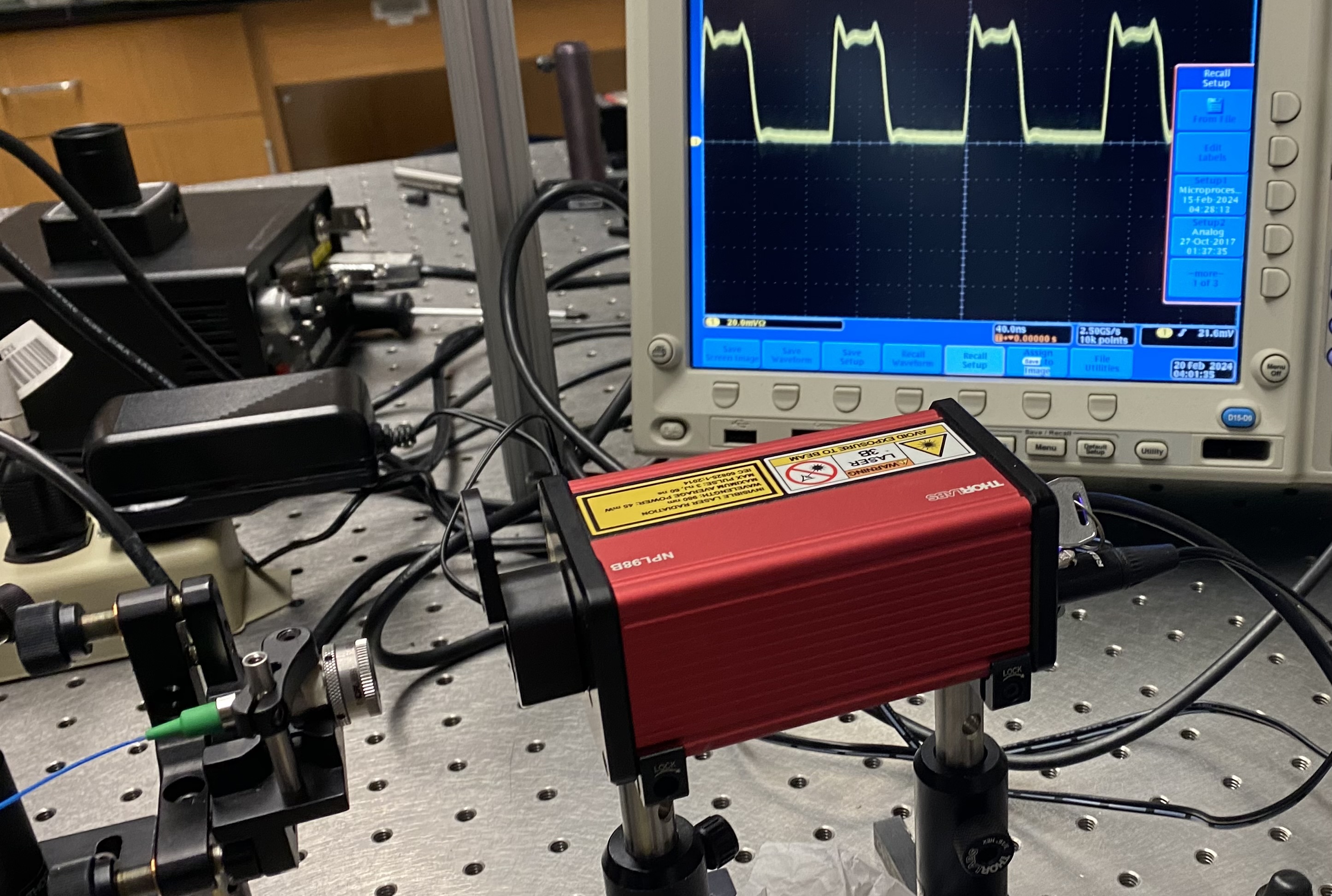}
    \caption{A photo of our nanosecond pulse laser, the fiber coupling interface, and the oscilloscope readout from our detector.}
    \label{setup}
\end{figure}

Finally, the output is passed into an inline polarizing beamsplitter, which splits the light into two channels based on the polarization state. One channel represents a final result of a 0 bit, and the other represents a 1 bit. The channels are then fed to two separate photon detectors and connected to an oscilloscope, which displays the power output from each channel. This output shows either a large signal in the 0-bit channel (detector 0), a large signal in the 1-bit channel (detector 1), or the optical power divided approximately equally between the two channels.  Bob’s detector results are displayed on an oscilloscope, where the readings for each channel can be displayed and uploaded to our computer for analysis. A diagram of our setup is shown in Fig. \ref{qkd_diagram}, and a photo of our laser and subsequent fiber-optic coupling is shown in Fig. \ref{setup}.

In our many-photon regime, where we maintain linear polarization, the polarization interactions can be described by Eq. (\ref{eq:malus}). If Bob chooses the same basis as Alice, he will rotate the polarization of light to the same measurement basis as the polarizing beamsplitter, leading to a large signal in one of the detectors. If he chooses incorrectly, the light will be rotated to a measurement basis 45 degrees offset from the polarizing beamsplitter, causing an even split between the signal in each detector. Although different, this is meant to represent the probabilistic 50/50 result that would happen to a single photon in a true BB84 implementation. Unlike the true single-photon implementation, the equal-power split directly reveals mismatched bases; this is a limitation of our demonstration.

The motorized polarization controllers work by applying a mechanical stress to the optical fiber, which induces birefringence, where there is a difference in refractive index based on the polarization axis. This causes the polarization components of the light to shift relative to one another, inducing a change in the overall polarization state. A specific servo-motor configuration corresponds to a certain applied stress and thus a certain shift in polarization, which is not necessarily limited to linearly polarized light. 

In order to calibrate the settings for these experiments, the separation used by the polarizing beamsplitter was used as a reference H/V linear polarization basis measurement. With the input guaranteed to be some linearly polarized state, we calibrated the states by manually maximizing the output of a given photon detector, adding additional wave plates as needed to switch from H/V to D/AD. Using Alice’s calibrated inputs, Bob’s desired polarization shifts were calibrated in a similar manner. In this way, we approximated the linearly polarized H/V and D/AD basis choices and measurements. Although this procedure is not exact, our many-photon system does not have to be as perfect as a true BB84 apparatus to demonstrate its principles. The random basis and bit choices for Alice and Bob were generated by the Arduino’s built-in pseudorandom generator, which is adequate for the educational purposes of this demonstration. 

\section{Results}

\begin{figure}[h]
    \centering
    \begin{subfigure}[b]{0.8\textwidth}
        \includegraphics[width=1.0\textwidth]{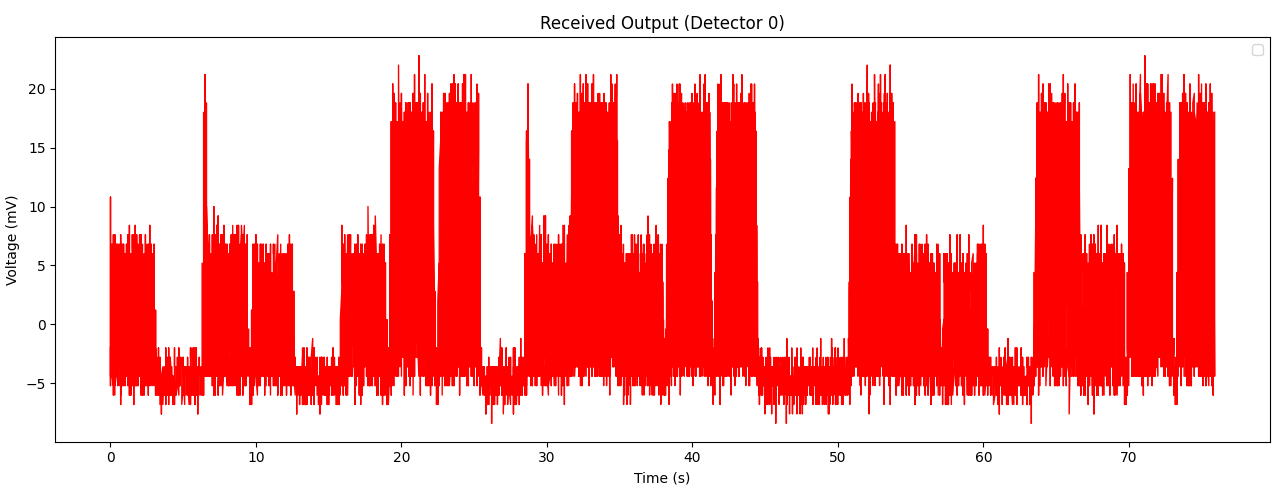}
        \caption{Oscilloscope reading for detector 0. A high voltage output corresponds to a received 0 bit, and a low voltage corresponds to a received 1 bit. Middle-level voltages correspond to mismatched bases.}
        \label{channel0}
    \end{subfigure}
    \begin{subfigure}[b]{0.88\textwidth}
        \includegraphics[width=0.91\textwidth]{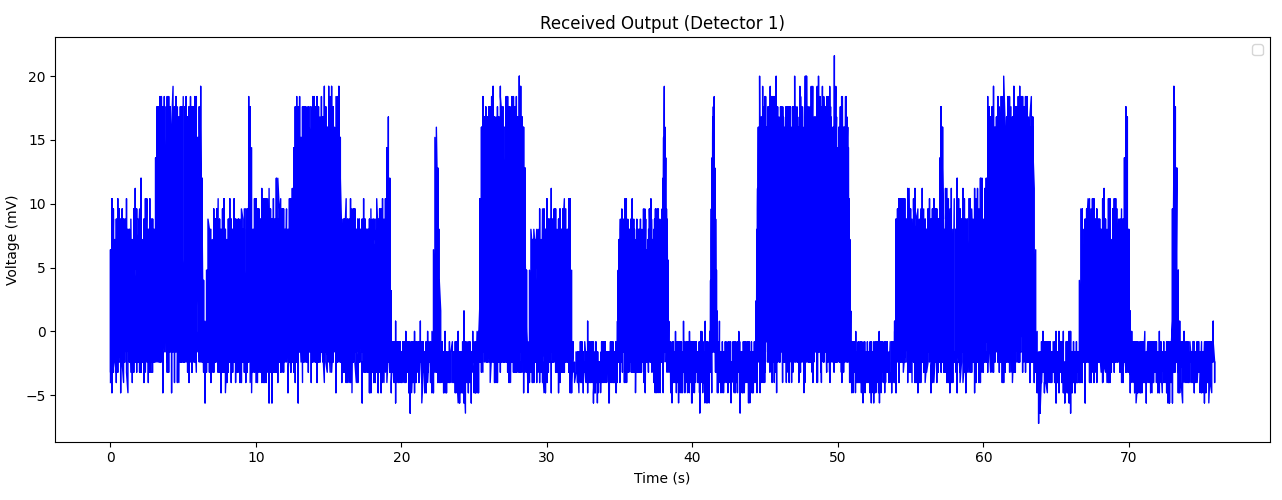}
        \caption{Oscilloscope reading for detector 1. Conversely, a high voltage corresponds to a received 1 bit, while a low voltage corresponds to a received 0 bit.}
        \label{channel1}
    \end{subfigure}
    \caption{This is the oscilloscope output from our 24-bit test case, separated by detector. Over time the signals change, as a different bit is sent from Alice to Bob. There is a short transition period between the transmissions, where the voltage fluctuates rapidly as the servo motors move. This period corresponds to the servo motors rotating the polarization to match the next transmission.}
    \label{channels}
\end{figure}

In order to test the functionality of our setup, we generated a random string of bits and transmitted it through our photonic system. After generating a random 24-bit string of 110110001001001101110100 and random bases for Alice and Bob, we transferred the key from Alice to Bob through the optomechanical setup, with photons collecting in two detectors. The data from these detectors were read out through an oscilloscope, which is shown in Fig. \ref{channels}. The desired outcomes are shown in Table \ref{tab:testcase}. This random 24 transmission sequence includes all eight possible combinations of bits (Alice) and bases (Alice and Bob), covering every possible polarization rotation in the protocol. It should be noted that this basis-sifting process does not correspond to the end of a real QKD implementation, and Appendix A briefly describes additional security measures taken after this step ensure a secure key.

\begin{table}[ht]
  \centering
  \caption{Test case details (h corresponds to the H/V basis, d corresponds to D/AD).}
  \label{tab:testcase}
  \begin{tabular}{@{} l >{\ttfamily}l @{}}
    \toprule
    \textbf{Field}          & \textbf{Value} \\
    \midrule
    Key                     & 110110001001001101110100 \\
    Alice’s bases           & dhhddddhhhhddhddhdhhdhhd   \\
    Bob’s bases             & hhdhdhdhhdhhdhddhhdhddhd  \\
    Matching indices (1-24)        & [2,\,5,\,7,\,8,\,9,\,11,\,13,\,14,\,15,\,16,\,17,\,20,\,21,\,23,\,24] \\
    Desired output          & 110010001101000           \\
    \bottomrule
  \end{tabular}
\end{table}

\begin{figure}[h]
    \centering
    \includegraphics[width=0.75\textwidth]{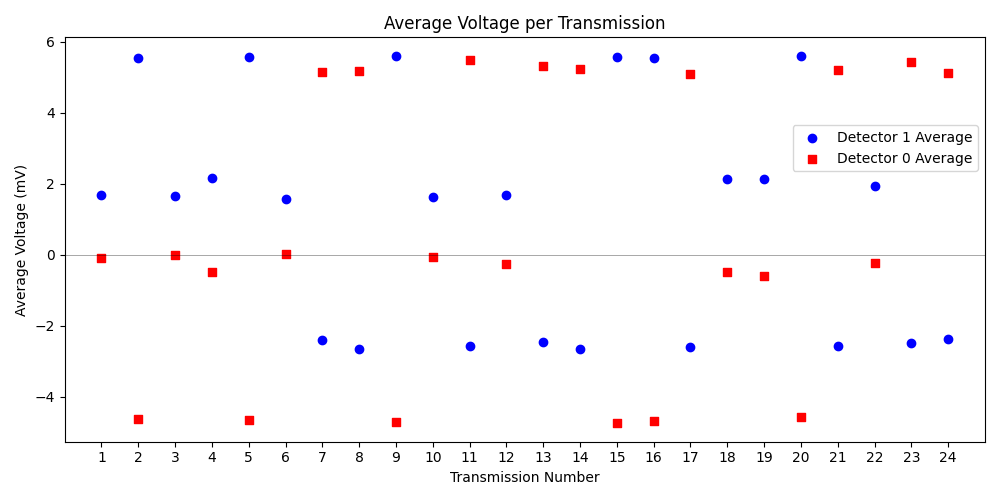}
    \caption{Average voltage of each transmission in both channels. Using the code for the polarization rotation apparatus to inform our averaging window, each transmission average used 2000 ms (200 included data points) with a 1200-ms (120 excluded data points) window reserved for servo motor transitions.}
    \label{results_diagram}
\end{figure}

On their own, the oscilloscope results visually suggest a functional protocol, as desired 0 bits result in a high detector 0 signal, desired 1 bits result in a high detector 1 signal, and mismatched bases result in a mixed signal. However, the evaluation process needs automation. In order to automate the process and develop quantifiable data, the oscilloscope results were averaged over time for each of the 24 transmissions. The resulting averages are plotted in Fig. \ref{results_diagram}.

The output shown in Fig. \ref{results_diagram} has paired readings of both detectors. Each point at which the voltage readout for neither channel is particularly high corresponds to an uncertain result, while pairs in which the channels are far apart are certain results. In this case, a large result in detector 1 corresponds to a 1 bit while a large result in detector 0 corresponds to a 0 bit. In order to quantify the consistency of these averages, Table \ref{tab:category_stats} categorizes each of the 24 pulse averages into three main categories based on the predicted output signal: bit 1 sent with matched bases, bit 0 sent with matched bases, and mismatched bases. The transmission averages in each category were then averaged, and the standard deviation was calculated to test consistency. 
 
\begin{figure}[h]
    \centering
    \includegraphics[width=0.75\textwidth]{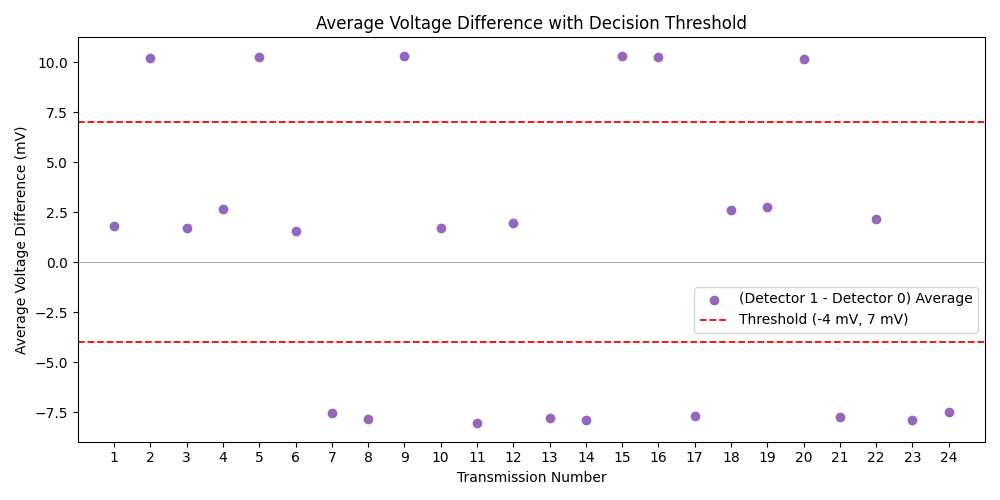}
    \caption{To maximize the reliability of the automated decision-making protocol, difference thresholds were chosen as far as possible from the voltages in each of the three categories. Above the threshold window, a 1 is chosen to be added to the key, and below the threshold window, a 0 is chosen to be added. Any values inside the window are assumed to be mismatched bases. Adapting the data from Table \ref{tab:category_stats}, the threshold window is over 10 standard deviations away from the closest category.}
    \label{threshold_diagram}
\end{figure}

Although the sample size is relatively small, Table \ref{tab:category_stats} indicates that each category is separated from its nearest neighbor by several tens of standard deviations. This points to a low risk of misidentifying an output and suggests a reliable protocol. Based on the voltage average for each transmission, a variety of thresholds would be able to cleanly separate each output category, correctly categorizing all 24 transmissions. Using the 24-bit average data, we created a voltage difference threshold to be used in future experiments, shown in Fig. \ref{threshold_diagram}. Although the threshold correctly characterizes each transmission, future trials to further test this threshold could help strengthen its validity.

\begin{table}[t]
\centering
\small
\caption{Detector output statistics by received signal category for the 24-bit trial. For each sent bit, an average of all detector data was taken. From these averages, the mean and standard
deviation are computed to determine the reliability of the output signal for each of the three possible output categories.}
\label{tab:category_stats}
\begin{tabular}{lcccc}
\toprule
\textbf{Category} & \textbf{$N$ (bits sent)} & \textbf{Detector 1 mean (mV)} & \textbf{Detector 0 mean (mV)} & \textbf{(Detector 1 - Detector 0) mean (mV)} \\
\midrule
Bit 1 sent (matched bases)   & 6 & $5.58 \pm 0.03$  & $-4.67 \pm 0.07$ & $10.24 \pm 0.06$\\
Bit 0 sent (matched bases)   & 9 & $-2.52 \pm 0.11$ & $5.25 \pm 0.14$ & $-7.77 \pm 0.18$\\
Mismatched bases             & 9 & $1.84 \pm 0.24$  & $-0.24 \pm 0.23$ & $2.09 \pm 0.46$\\
\bottomrule
\end{tabular}
\end{table}

Using this threshold, the mismatched bases are eliminated when Alice and Bob compare their results, producing the key 110010001101000. This is an exact match to the desired output key. Using the method of 2000-ms averaging windows and 1200-ms transition periods, combined with the thresholds shown in Fig. \ref{threshold_diagram}, one could automate this process to generate longer keys than the one shown above.

\section{Discussion}

Our setup shows a simple, accessible way to emulate the process of quantum key distribution in classical optics. We are able to demonstrate the encoding, transmitting, and receiving aspects of the BB84 protocol, as described in Appendix A. In an ideal single-photon protocol, an eavesdropper who intercepted and re-sent the states would introduce errors that Alice and Bob would detect when they disclosed and compared a sample of their sifted bits. In our many-photon apparatus, however, this detection step is only partial, as discussed below.

Our apparatus uses a large number of photons per pulse, rather than a single photon carrying one quantum state. This makes the protocol easier to visualize, as mismatched bases would produce an approximately even split of optical power between the two detector channels. This also makes our protocol easier to implement, as there is room for error during our calibration of the apparatus. In the detection step, Alice and Bob compare their results to generate a final key, but our protocol does not implement the single-photon security mechanism of BB84, which is elaborated on in Appendix A. Each pulse contains many copies of photons in the same polarization state; thus the no-cloning theorem does not prevent an eavesdropper from intercepting a small fraction of the pulse and measuring it, leaving the majority of photons undisturbed. An important future direction would therefore be to move toward the single-photon regime using weak coherent pulses \cite{Lo2005Decoy}, which would enable this protocol to detect an eavesdropper. Introducing an eavesdropper into the system would also be a valuable demonstration in the weak coherent pulse case, in the form of a separate photon detector and source.

Our results suggest a reliable protocol with relatively low variation in results, although a future study with a wider variety of transmitted keys and longer bitstrings could better characterize the accuracy of this implementation. It is also important to note that negative voltages appear in our results, which would not be physically possible if our detectors were perfectly calibrated with zero voltage corresponding to zero intensity. In Appendix B, we included the Thorlabs specifications for the photon detectors, including the dark offset in Table \ref{tab:acquisition_settings}. We believe this fixed, detector-specific offset is causing our negative voltages. This offset does not inhibit our ability to distinguish different results relative to our threshold; it simply imparts a vertical shift into our raw data.

\section{Conclusion}
Our apparatus reproduced the expected distilled key for the transmitted test sequence, demonstrating the polarization-encoding and basis-sifting steps of BB84 for our optomechanical setup. Our many-photon system demonstrates the polarization optics and logical structure of the protocol without the precision required for a truly secure key distribution. We were able to properly represent the choice of bits and bases, storing those data in the polarization states of photons. Based on our Arduino code for the servo motors, we were also able to develop an automated procedure for measuring the result of each transmission, calculating averages from the time-domain oscilloscope data. Using this procedure for our 24-bit test case, the three output categories were clearly distinguishable. 

Our photonic implementation mimics real-world fiber-optic communication. Rather than sending a preexisting secret through the channel, the protocol is designed to distill a shared random key from Alice’s and Bob’s correlated measurements. Once such a key has been established and verified, it can be used in the encryption of future messages.

As quantum computers become more powerful, the need for secure data encryption only grows. As new methods of QKD and quantum-resistant encryption are developed\cite{AdvancesQKD,ChhetriPQCSurvey}, education on these techniques becomes more important. Quantum key distribution is far from the only quantum computing cybersecurity protocol, and fields such as post-quantum cryptography (PQC) are working to solve the same problem. 

Our specific protocol is designed for use in educational settings, due to our equipment and the size of our system. We offer an emulation of BB84 that is easy to customize, program, build, and observe, in addition to being relatively cost-effective. It requires knowledge of optics as well as basic principles of computer science and electronics. In addition to these areas, it can be used as a starting point for students to learn and experiment with quantum security, a field which is constantly evolving and gaining relevance.

\section{Acknowledgments}
We would like to thank Mark Hannum for introducing us to quantum computing and quantum security through the quantum program at Thomas Jefferson High School for Science and Technology, as well as giving us inspiration and guidance throughout the completion of this project.

\bibliographystyle{aipnum4-2}
\bibliography{revisedrefs}

@inproceedings{Bennett1984,
   author       = "C. H. Bennett and G. Brassard",
   year         = "1984",
   title        = "Quantum Cryptography: Public Key Distribution and Coin Tossing",
   booktitle    = "Proceedings of the {IEEE} International Conference on Computers, Systems and Signal Processing",
   pages        = "175--179",
   address      = "Bangalore, India",
   publisher    = "IEEE",
}

@incollection{NoCloningTheorem,
   author       = "F. A. Bais and J. D. Farmer",
   year         = "2008",
   title        = "The Physics of Information",
   booktitle    = "Philosophy of Information",
   editor       = "P. Adriaans and J. {van Benthem}",
   series       = "Handbook of the Philosophy of Science",
   publisher    = "North-Holland",
   address      = "Amsterdam",
   pages        = "609--683",
   doi          = "10.1016/B978-0-444-51726-5.50020-0",
}

@book{NumberFieldSieve,
   editor       = "A. K. Lenstra and H. W. Lenstra",
   year         = "2006",
   title        = "The Development of the Number Field Sieve",
   publisher    = "Springer",
   address      = "Berlin, Heidelberg",
   doi          = "10.1007/BFb0091534",
}

@article{ShorsPaper,
   author       = "P. W. Shor",
   year         = "1997",
   title        = "Polynomial-Time Algorithms for Prime Factorization and Discrete Logarithms on a Quantum Computer",
   journal      = "SIAM J.\ Comput.",
   volume       = "26",
   number       = "5",
   pages        = "1484--1509",
   doi          = "10.1137/S0097539795293172",
}

@unpublished{QubitsToBreakRSA,
   author       = "C. Gidney",
   year         = "2025",
   month        = "22~" # may,
   title        = "How to Factor 2048 bit {RSA} Integers with Less Than a Million Noisy Qubits",
   eprint       = "2505.15917",
   archivePrefix= "arXiv",
   url          = "https://arxiv.org/abs/2505.15917",
}

@article{AdvancesQKD,
   author       = "V. Zapatero and T. {van Leent} and R. Arnon-Friedman and W.-Z. Liu and Q. Zhang and H. Weinfurter and M. Curty",
   year         = "2023",
   month        = "2~" # feb,
   title        = "Advances in Device-Independent Quantum Key Distribution",
   journal      = "npj Quantum Inf.",
   volume       = "9",
   number       = "1",
   doi          = "10.1038/s41534-023-00684-x",
}

@unpublished{DiscreteLogarithms,
   author       = "M. Eker{\aa}",
   year         = "2024",
   month        = "9~" # sep,
   title        = "Revisiting Shor's Quantum Algorithm for Computing General Discrete Logarithms",
   howpublished = "arXiv:1905.09084",
   url          = "https://arxiv.org/abs/1905.09084",
}

@article{EllipticCurves,
   author       = "J. Proos and C. Zalka",
   year         = "2003",
   title        = "Shor's Discrete Logarithm Quantum Algorithm for Elliptic Curves",
   journal      = "Quantum Inf.\ Comput.",
   volume       = "3",
   number       = "4",
   pages        = "317--344",
}

@inproceedings{QKDOverview,
   author       = "A. I. Nurhadi and N. R. Syambas",
   year         = "2018",
   month        = "13~" # jul,
   title        = "Quantum Key Distribution ({QKD}) Protocols: A Survey",
   organization = "IEEE",
}

@article{Lo2005Decoy,
   author       = "H.-K. Lo and X. Ma and K. Chen",
   year         = "2005",
   title        = "Decoy State Quantum Key Distribution",
   journal      = "Phys.\ Rev.\ Lett.",
   volume       = "94",
   number       = "23",
   pages        = "230504",
   doi          = "10.1103/PhysRevLett.94.230504",
   publisher    = "American Physical Society",
}

@unpublished{ChhetriPQCSurvey,
   author       = "G. Chhetri and S. Somvanshi and P. Hebli and S. Brotee and S. Das",
   year         = "2025",
   month        = "12~" # oct,
   title        = "Post-Quantum Cryptography and Quantum-Safe Security: A Comprehensive Survey",
   eprint       = "2510.10436",
   archivePrefix= "arXiv",
   url          = "https://arxiv.org/html/2510.10436v1",
}

@article{SharmaSaxena2025,
   author       = "N. Sharma and V. Saxena",
   year         = "2025",
   title        = "A Comparative Analysis of Error Correction Techniques Used for {QKD} Protocols",
   journal      = "J.\ Opt.",
   pages        = "1--7",
   doi          = "10.1007/s12596-025-02818-0",
}

@incollection{PavlovicSeidel2025,
   author       = "D. Pavlovic and P.-M. Seidel",
   year         = "2025",
   title        = "Communication Channels, Protocols, and Authentication",
   booktitle    = "Introduction to Security Science: Basic Concepts and Mathematical Foundations",
   series       = "Information Security and Cryptography",
   publisher    = "Springer",
   address      = "Cham",
   pages        = "83--106",
   doi          = "10.1007/978-3-032-00949-4_7",
}

\clearpage
\appendix

\section{Appendix A: BB84 Algorithm Steps}
\label{sec:algorithmsteps}

\subsubsection{\label{sec:level3}Phase 0: Encoding}

Alice and Bob agree to an encoding table. When bits and bases are selected, they are converted into quantum states based on this table. We use the earlier concept of H/V and D/AD bases in order to establish measurement in our photonic setting. We can represent both a 0 bit and a 1 bit with either basis. The horizontal and diagonal basis vectors represent a 0 bit, while the vertical and antidiagonal basis vectors represent a 1 bit.

\begin{table}[h!]
    \centering
    \caption{The encoding between bases and the binary representation of the data.}
    \label{encoding}
    \renewcommand{\arraystretch}{1.2}
    \begin{tabular}{|c|c|c|}
        \hline
        \textbf{Binary} & \textbf{H/V-Basis} & \textbf{D/AD-Basis} \\
        \hline\hline
        0 bit & H & D \\
        1 bit & V & AD \\
        \hline
    \end{tabular}
\end{table}

\subsubsection{\label{sec:level3}Phase 1: Sending}

Alice then selects a random sequence of bits, followed by a random sequence of bases. She then encodes the bits using the bases chosen.

\begin{table}[h!]
\centering
\caption{Alice encoding a bitstring of 01100 into quantum states.}
\label{sending}
\renewcommand{\arraystretch}{1.2}
    \begin{tabular}{|c||c|c|c|c|c|}
        \hline
        \textbf{Alice} & \textbf{1} & \textbf{2} & \textbf{3} & \textbf{4} & \textbf{5} \\
        \hline\hline
        Bits & 0 & 1 & 1 & 0 & 0 \\
        Bases & H/V & H/V & D/AD & H/V & D/AD \\
        States & H & V & AD & H & D \\
        \hline
    \end{tabular}
\end{table}

\subsubsection{\label{sec:level3}Phase 2: Receiving}

Bob receives only Alice’s encoded optical state, which contains information about both the bit and basis. Bob then independently chooses a random sequence of bases (each one either H/V or D/AD) to measure the resulting polarization. Next, Bob measures Alice’s photons with the chosen bases. Finally, Bob decodes the quantum states according to Table \ref{receiving}.

In this ideal case, measurement of an H or V state in the H/V basis has a 100\% accuracy, the same as measuring a D or AD state in the D/AD basis. However, measuring a basis vector in the other basis creates a 50\% chance for each of the two possible outputs, as the H/V basis vectors are exactly 45 degrees apart from the D/AD basis vectors, as shown in Fig. \ref{phase_diagram}. Thus, if Alice and Bob choose the same basis, Alice’s original vector is unchanged and the bit stays the same. However, if Alice and Bob choose different bases, Bob’s measurement projects Alice’s state onto one of his two basis vectors, each with probability one-half, so the decoded bit has exactly a 50\% chance of matching Alice’s. Before this measurement, Alice’s state is definite in her own basis; the randomness appears only through Bob’s projection onto a different basis.

\begin{table}[h!]
    \centering
    \caption{The possible measurements Bob can make on Alice's transmitted states, followed by the decoded bits.}
    \label{receiving}
    \renewcommand{\arraystretch}{1.2}
    \begin{tabular}{|c||c|c|c|c|c|}
        \hline
        \textbf{Bob} & \textbf{1} & \textbf{2} & \textbf{3} & \textbf{4} & \textbf{5} \\
        \hline\hline
        Alice state & H & V & AD & H & D \\
        Bob bases & D/AD & H/V & H/V & H/V & D/AD \\
        Possible measurements & D or AD & V & H or V & H & D \\
        Measured state & D & V & H & H & D \\
        Decoded state & 0 & 1 & 0 & 0 & 0 \\
    \hline
    \end{tabular}
\end{table}

\subsubsection{\label{sec:level3}Phase 3: Comparing}

Once Bob records the final bits, Alice and Bob then publicly compare only their bases, keeping their measured bits private. If Alice and Bob used the same basis for a given bit, then they keep that bit in their final key. However, if they used different bases, then they throw out the result, due to the uncertainty mentioned above. Thus, the resulting key only consists of bits that stayed the same from Alice to Bob, due to their choosing of the same basis. 

In a full QKD system, several further steps follow basis sifting. Alice and Bob publicly disclose a small portion of their bits and discard the result if their error rate is abnormal, which would suggest the presence of an eavesdropper or a faulty channel. Even in a secure channel, there is the chance of noise, so Alice and Bob must perform error correction between their bits. Assuming a relatively small error rate, they can reveal a small amount of information in order to correct their bitstrings, such as with a parity check\cite{SharmaSaxena2025}. To minimize the information that a potential eavesdropper may have, they can also hash their bits through a function which would take multiple input bits to predict any single output bit. This reduces the information that an eavesdropper with a small amount of stolen bits has about the secret key. Another important caveat of QKD is a man-in-the-middle attack, where Eve pretends to be Alice to Bob and pretends to be Bob to Alice. In order to avoid this, the communication channel between Alice and Bob has to be authenticated, which is a separate problem that must be addressed beforehand\cite{PavlovicSeidel2025}. Importantly, this algorithm only distributes a key; it does not automatically encrypt a message. If a conventional cipher is used to encrypt messages, the security of the communication is now dependent on the security of the cipher. Our apparatus emulates the encoding and sifting steps but does not implement this postprocessing.

\begin{table}[h!]
    \centering
    \caption{The final comparison, producing a key of 100.}
    \label{comparing}
    \renewcommand{\arraystretch}{1.2}
    \begin{tabular}{|c||c|c|c|c|c|}
        \hline
        \textbf{Compare} & \textbf{1} & \textbf{2} & \textbf{3} & \textbf{4} & \textbf{5} \\
        \hline\hline
        Alice bases & H/V & H/V & D/AD & H/V & D/AD \\
        Bob bases & D/AD & H/V & H/V & H/V & D/AD \\
        Alice bits & 0 & 1 & 1 & 0 & 0 \\
        Bob bits & 0 & 1 & 0 & 0 & 0 \\
        \hline
        Final key & & 1 & & 0 & 0\\
        \hline
    \end{tabular}
\end{table}

\section{Appendix B: Materials}
\label{sec:materials}

All optical components were purchased from Thorlabs. The part numbers are listed as enumerated in Fig. \ref{qkd_diagram}, along with the part names below. Table \ref{tab:acquisition_settings} shows the acquisition settings used to generate the results, including the dark offset of the PDA10A2 detectors, which we believe to be the reason for the negative voltages.

\begin{enumerate}
    \item \textbf{Laser:}  Nanosecond Pulsed Laser Diode System, 980 nm, 20-ns pulse. (NPL98B)
    \item \textbf{Beam collimator:} Adjustable Fiber Collimator, FC/APC, f = 2.0 mm, 600-1050 nm AR coating. (CFC2A-B)
    \item \textbf{Polarization maintaining (PM) fiber-optic cables:} PM Patch Cable, PANDA, 980 nm, Ø3 mm Jacket, FC/APC, 2 m long. (P3-980PM-FC-2)
    \item \textbf{Inline polarizer:} 	In-Line Fiber Polarizer, 980 ± 10 nm, PM/PM Pigtail, FC/APC. (ILP980PM-APC)
    \item \textbf{Dynamic polarization rotator apparatus:} Motorized Fiber Polarization Controller for Ø900-µm Jacket Fiber, two paddles, Ø18 mm. (MPC220)
    \item \textbf{Single-mode (SM) fiber-optic cable:} Single Mode Patch Cable, 980-1650 nm, FC/APC, Ø900-µm Jacket, 2-m long. (P3-1064Y-FC)
    \item \textbf{Inline polarizing beamsplitter:} Fused Fiber Polarization Combiner/Splitter, 980 $\pm$ 15 nm, FC/APC. (PFC980A)
    \item \textbf{Connector between fiber and detector:} FC/APC Fiber Adapter Cap with Internal SM1 (1.035"-40) Threads, Narrow Key (2.0 mm). (S120-APC2)
    \item \textbf{Photon detector:} Si Fixed Gain Detector, 200–1100 nm, 150 MHz BW, 0.8 mm\(^{2}\), universal 8-32/M4 taps. (PDA10A2)
    
    \item \textbf{Connector between fibers:} 	FC/APC Single Mode Connector, Ø126-µm Bore, Ceramic Ferrule, Ø900-µm boot. (30126A9)
    
\end{enumerate}

\begin{table}[t]
\centering
\caption{Oscilloscope settings and detector specifications.}
\label{tab:acquisition_settings}
\begin{tabular}{ll}
\toprule
\textbf{Parameter} & \textbf{Value} \\
\midrule
Oscilloscope model & Tektronix MSO4032, Mixed Signal Oscilloscope \\
Sample rate & 100 Sam/s \\
Voltage increment & 0.8~mV  \\
Probe attenuation & $\times 1$ \\
\midrule
Detector model & Thorlabs PDA10A2, Si Fixed Gain Detector\\
Detector gain & 10~kV/A \\
Dark offset & $\pm 10$~mV \\
\bottomrule
\end{tabular}
\end{table}

All code used for our implementation can be found in our GitHub repository: \url{https://github.com/karthikcsq/QKD_Protocol}.
\end{document}